\documentclass[12pt]{article}
\usepackage{amsmath}
\usepackage{graphicx,epsf}
\usepackage{enumerate}
\usepackage{natbib}
\usepackage{url} 

\newcommand{\blind}{0}

\begin{document}
\def\spacingset#1{\renewcommand{\baselinestretch}%
{#1}\small\normalsize} \spacingset{1}


\if0\blind
{
  \title{\bf George Udny Yule and the Interpretation of Regression Betas}
  \author{Francesco Corielli\hspace{.2cm}\\
    Department of Finance and IGIER, Bocconi University of Milan\\
 }
  \maketitle
} \fi

\if1\blind
{
  \bigskip
  \bigskip
  \bigskip
  \begin{center}
    {\LARGE\bf Title}
\end{center}
  \medskip
} \fi

\bigskip
\begin{abstract}
Initially applied in astronomy and geodesy, the linear regression model aimed to find the best estimates for parameters with predefined meanings (e.g., orbital elements, geodetic constants). As its use expanded to other disciplines—often to summarize data without an underlying theoretical model—the need for a general interpretation of regression betas arose. Early attempts by Galton and Karl Pearson met with mixed success. G. U. Yule was the first to develop a general statistical interpretation, the culmination of efforts begun in 1896. Yule's interpretation is based on the partial regression theorem, which he proved in 1907. 
\end{abstract}

\noindent%
\vfill

\newpage
\spacingset{1.45} 

\section*{Introduction.}

The linear regression model is ubiquitous across diverse scientific fields today. Often, the regression betas in the model correspond to quantities whose meaning is already established by a predefined theory, model, or measurement system. In such cases, these betas require no further interpretation beyond what the underlying theory dictates; the linear regression model simply aims to provide sound statistical estimates of these parameters.

However, in many  fields the linear regression model is used in order to summarize relationships between observed quantities in a setting where no precise a priori theory exists. In this case, almost by necessity, some ``interpretation'' of the meaning of the regression beta is required.

The original setting of the linear model is the first: in its late XVIII century birth and first maturity in the works of legendre and Gauss, the model was applied to the estimation of orbital parameters and geodesy problems. Betas corresponded to orbital elements or to geodesy measures. No need for further interpretation was required. What was done, mostly by Gauss, was to define and study  hypotheses on measurement errors under which least squares estimates were optimal.

Starting with the work of Galton, then of Pearson, Yule and many others, the linear regression model began to be applied in different fields: first genetics, then economics and more in general the social sciences. In these fields no or just partial formal theoretical models were available and the regression betas were usually understood as measures of correlations between observables. Usually, not always, we'll see how the original interpretation by Galton was on a different line.

In these fields, linear regression was used to answer statistical questions, most frequently of a predictive kind, as, for instance, the correlation of offspring characters with parental characters or the effect of a relief policy on pauperism. In such fields, 
betas, no more corresponding to pre defined theoretical constructs, required some new interpretation.

The object of this paper is to tell the tale of how the modern way to interpret linear regression betas was born between the end of the XIX and the beginning of the XX century, over the span of almost forty years. How it started with some intuition by Galton. Went thru a more detailed but still incomplete analysis of Karl Pearson and was completed by Yule in a series of paper which, incidentally, contain one of the first examples of what today we call econometrics.

It may be useful to begin from the end of the story: with an example of how the meaning of the regression beta is explained today to students. The example comes from  a quite popular textbook of Econometrics:  Gujaraty (2009, p. 191, section 7.3) 
\begin{quotation}
The meaning of partial regression coefficient is as follows: $\beta_2$
measures the change in the mean value of $Y,E(Y)$, per unit change
in $X_2$, holding the value of $X_3$ constant. 
Put differently,
it gives the "direct" or the "net"
effect of a unit change in $X_2$ on the mean value of $Y$, net
of any effect that $X_3$ may have on mean $Y$
\end{quotation}
This statement, echoing many in pedagogical literature, prompts a crucial question:  how is it posible to ``hold the value of $X_{3}$ constant''? And: what has this to do with the `` "direct" or  "net"
effect of a unit change in $X_{2}$ on the mean value of $Y$''.

As the very next paragraph in Gujaraty (2009) asks, "How do we actually go about holding the influence of a regressor constant?" The answer, which follows, lies in a detailed analysis of linear regression based on the partial regression theorem, first proven by Yule in 1907 for this very purpose. This theorem marks a critical point in our origin story.

The section ends with:
\begin{quotation}
Do we have to go through this multistep procedure every time we want to find out the
true partial regression coefficient? Fortunately, we do not have to do that, for the same job
can be accomplished fairly quickly and routinely by the OLS procedure discussed in the
next section. The multistep procedure just outlined is merely for pedagogic purposes to
drive home the meaning of “partial” regression coefficient.
\end{quotation}

In what follows, we shall see how, actually, the passage from some unspecified idea of ``keeping the rest constant'' to a precise definition of ``controlling for'', which today requires few paragraphs, was not easy and immediate. Even if we abstract, which would be wrong, from the work of Galton and Karl Pearson which completely renovated and enlarged the application span of the linear model, it took Yule from 1896 when, de facto, for the first time he mentioned the ``keep the rest constant''clause,  to 1907 when, presenting what today we call the partial regression theorem, he clearly stated, with a stern style quite infrequent in his, usually subdued, scientific prose, the importance of the result for the interpretation of a linear regression.

We could argue that the historical difficulty in passing from an intuitive idea about ``keeping the rest constant'' to a formal definition of ``controlling for'' is today mimicked by the paths followed in teaching and learning regression. As discussed, for instance, in the celebrated chapter 13 of Mosteller and Tukey (1977), failure of fully understanding this point is the origin of many problems in interpreting regression coefficients still common nowadays.

\subsection*{Structure of the paper.}

This paper is organized as follows.

The first section sets the stage
by providing an overview of the evolution of regression betas interpretation
from late 18th century to Galton's discovery of the "regression
effect".

The second section is dedicated to some notation and a
short summary of the Partial Regression Theorem (PRT).

The third, and main section, shall summarize the
history of regression interpretation with main focus on the works of Pearson and
Yule.


A last section is dedicated to few  suggestions
we think useful for teaching regression.

\section{From theory of errors to regression as best fit.}

To fully appreciate Yule's contribution, we must trace the evolution of linear models from the mid-18th century "theory of errors." This intellectual journey led to the discovery of least squares and its initial applications in astronomy and geodesy. During the latter half of the 19th century, these methods, now under the new name of regression, expanded beyond their original domains, finding application first in the biological sciences, and subsequently in the social sciences.

The story of the
theory of errors and least squares has already, masterfully, been
written in detail and we refer for this to Stigler (1981), Seal (1967)
and Aldrich (1998). 

In fact, our attempt is just that of adding some new and, we think, important
consideration to this literature.

For the sake of brevity,  we shall recur to anachronisms,
e.g. using the term "linear regression'' for
something that received this name only beginning with Galton in the last 30 years of the 19th century

Linear regression and least squares originate from a problem which became to be discussed  toward the end of 17th century and can be summarized
as follows.

The measure of a quantity (a "parameter'')
$\theta$, e.g. an orbital element, the length of a meridian, is required.
This quantity can only be observed indirectly, and each observation
gives, in general, a different result. 
Since $\theta$ is assumed constant, we are led to the idea that the
variability of results is due to "observation errors''. The question
is how to use the data in such a way to get the "best estimate''
of $\theta$. 
This requires a "theory of errors'': a description of how observation
errors affect measures of $\theta.$

This may seem obvious, today, but it was not always the case when
this problem was first considered. 
In Stigler (1986, p. 94), for instance, we see the critique made by Thomas Bayes,
in a letter to John Carlton, to an idea of Thomas Simpson's: taking
the average of many measures should reduce measurement errors. Bayes
corrects Simpson: what we actually get is just a mean of errors, this
is not 0 in general. For instance, it is not zero if errors are systematic.
Hypotheses are required. Simpson incorporated
Bayes observation (without quoting) in a new version of his work.
At the time of Simpson and Bayes, averages where already used to summarize
data. What was novel,
was the use of average as "estimator'' of an underlying parameter.
Bayes understood that assumptions are necessary for this to work.  

A similar problem arose in the fields of Astronomy
and Geodesy.
The objective was that of estimating a vector of
$k+1$ parameters $\theta$ (e.g.: orbital elements for comets and
asteroids, length of meridians) on the basis of $n\times k+1$ observables
$W$ (e.g. positions of a celestial object at given times). The observables
$W$ and parameters $\theta$, had to satisfy some set of "laws''
(Kepler's laws for orbits, geometry of curved surfaces for Geodesy).
These could be written, maybe after linearization and coordinate change,
as $W\theta=0$.
Due to approximation, observation error etc., errors $u$ had to be
admitted: $W\theta=u$. 
Usually, $W\theta$ was written as $Y-X\beta$, where $Y=W_{j}$ for
some $j$, $X=W_{-j}$ and $\beta_{i}=\theta_{i}/\theta_{j}$ (suppose
$\theta_{j}\ne0$). We then have $Y-X\beta=\epsilon$, commonly written
as $Y=X\beta+\epsilon$ where $\epsilon=u/\theta_{j}$.

Researchers were in need of a "theory of errors'' to be applied
to $\epsilon$ and a way to estimate $\beta$.
Legendre and Gauss dealt with this problem and found the least squares
solution at the beginning of 19th century. 
There exists a strong analogy between what Legendre and Gauss did and
the Simpson-Bayes discussion 50 years before.
Legendre's approach (Legendre (1805) p. viii, and Appendice p. 72)
mirrored Thomas Simpson's. Once specified the equations describing observed comets positions  as functions Kepler's orbital elements, least squares aimed to find
the values of the orbital elements (parameters) which minimize the
sum of squared errors. The "fit to the data'' is maximized and
this is assumed to be the best possible solution. Not only in terms of fit
to the data, but also in terms of the quality of the estimate of $\beta$.

Crucially Legendre, like Simpson, does not discuss or characterize in any precise way hypotheses
on $\epsilon$ which would justify the use of the method to estimate
$\beta$.

Gauss, with an attitude similar to Bayes', was much more cautious. He recognized that simply maximizing the fit to observations
and minimize didn't necessarily lead to a good estimate of $\beta$. 

In fact, at the very beginning of Gauss (1821, translation by Stewart (1995)), Gauss discusses
errors.\footnote{Gauss (1821, translation 1995) is a  theory work. Two examples from Geodesy are in the "Supplementum". Gauss used least squares in astronomical problems  at the end of 18th century.} He distinguishes errors which today would be called random
measurement errors: zero expected value, uncorrelated with the conditions of measure and
with suitably stable distribution, and errors which today would be
called systematic errors. Gauss restricts his analysis to the first
kind of errors. Under this hypothesis, Gauss shows that maximizing
fit (minimize the sum of squared errors) also yields a "good estimate''
of $\beta$. Under some further hypothesis, this, actually, is the best possible estimate.
This is what we today call the ``Gauss Markov'' theorem.

The success of least squares and linear models in the fields of Astronomy
and Geodesy was quick and widespread.
Not surprising, in these fields no need arose for any interpretation
of $\beta$-s. This was a set of parameters in a well defined "structural" model
(e.g. orbital elements whose relations satisfy Kepler's laws) and needed no further interpretation.

Also, it is not surprising if, given its success, the model spread
to other fields. Fields where no precise theory was available, only
data waiting to be optimally summarized. 
In the first half of the 19th century we already see examples in this
direction. The work of Quetelet is an important case and we refer
for this to Stigler (1986) ch. 5.

With no modeling assumption, except linearity and existence of the
required expected values, we can compute the regression $E(Y|X)=X\gamma$ and we
know that $E(Y-X\gamma|X)=0$. Which implies $E(X'X)\beta=E(X'Y)$
the moment counterpart to least squares normal equations from which
$\gamma$ can be evaluated. Here $X\gamma$ does not correspond to
a "model'' it is just a tool to get a best fit or best forecast of $Y$.

By comparison, given the theory + noise model $Y=X\beta+\epsilon$,
Gauss basic assumption is that $E(\epsilon|X)=0$ so that $X\beta$
is  a regression function. This implies that $\gamma=\beta$ and, under added hypotheses, 
the methods that best fits the data (regression) also yields the best estimate of $\beta$ (the parameters of a theory). 

If we cannot suppose $E(\epsilon|X)=0$, our linear model is no more
a linear regression, $\beta\ne\gamma$. The regression $\gamma$
is still useful for computing a "best fit'' of $Y$ but is not useful for evaluating
$\beta$. (This is nothing but Bayes critique of Simpson).


Gauss goes beyond Legendre because he does show which hypotheses are
necessary in order to transform a "best fit to $Y$" tool (regression) into
a tool to estimate "structural parameters''.

Under these hypotheses, the regression parameter
$\gamma$ shares the interpretation of the model parameter $\beta$.

When linear regression begun to be used in settings where there was
no precise "theory'' model and no predefined meaning for $\beta$, no problem arose in computing $\gamma$
and get a best fit, but a new legitimate question arose about the
interpretation of $\gamma$.

In order to be useful in all fields where regression is used, the
answer to this question should be general and independent on the specific
context. It should only depend on the properties of the "best fit''
method itself. 

This general interpretation is presented for the first time in Yule (1907). This is the endpoint of a sequence of
 attempts by Galton, Pearson and Yule himself.

Both Yule and Karl Pearson were motivated by Galton's work on heredity. To complete our stage setting we need to briefly summarize this. 

In a series of empirical studies about the weight of seeds in several
generations of sweet peas and about anthropometric measures for parents
and offspring, mainly but not exclusively, stature, Galton realized
that deviation from the average of a character in the offspring was
"correlated'' with the same deviation in parents. On average, if
parental characters in excess of the mean were grouped by class of
values, offspring characters in excess of the mean were of the same
sign but smaller size.

After some change in terminology Galton called this empirical phenomenon
"regression'' (to the mean).
Galton also observed that, with good approximation, this regression
was constant for all classes of values of parents' character in excess.
Regression was, then, at least approximately linear and Galton's constant, became the ''regression'' parameter. 

Galton did not use least squares, but a set of graphical and quantile
based ad hoc techniques. The connection with least squares was made
precise by others. 

Regression, as observed by Galton, is a statistical fact justified
by its success in fitting the data. 
No precise heredity model did exist in order to give an interpretation
of the regression  fact or of the regression parameter. Galton does not "estimate'' a
predefined parameter in a physical or geometrical model from noisy
observations as Gauss did.

No surprise if Galton wondered about the meaning of the regression
parameter.
In the absence of a precise heredity model, Galton interpreted the regression parameter as a "fraction" of the character's deviation from the mean in parents that was "passed to" the offspring. 

It would be fascinating to delve into the reasons behind this "fractional heredity" choice and its connection to Charles Darwin's initial belief in pangenesis theory (a belief Darwin later abandoned), such a discussion would require a dedicated paper.\footnote{For more on pangenesys, which Galton, unwillingly, partially disproved in experiments
he led in 1869-71 see Browne (2002) ch. 8.}
What is relevant for us  is that Galton held that a measurable character of
offspring was function, thru some unknown mechanism, of the average
of the same character of parents and this function was well summarized
by a single parameter. 

This assumption  had consequences in Galton's analysis and, in practice, impeded Galton's development of regression to the multivariate setting. In fact, when defining "parents'
stature'' Galton did record maternal and paternal stature but did
not use them separately. He first converted maternal to paternal stature
by multiplying it by a constant greater than one and, then, used as
variable average of the paternal stature and the corrected maternal stature.

While Galton contemplated regressions with many independent
variables, he seems never to have dealt with this in detail (Pearson
(1930), pag. 78). 


 Karl Pearson's attempts to extend Galton's analysis to the case of two $X$s (Pearson (1896)).
The regressors considered by Pearson are two separate characters of different parents .
Pearson tries to keep Galton's interpretation in the new setting.
While the result is,  as we'll see, not fully satisfactory, Pearson does
not feel the need to  progress further toward a more general interpretation.

Yule found himself in a similar quandary when he applied multiple
regression to another "theory-less'' problem: the connection of
pauperism with aid in and out of workhouses (Yule (1897), (1899)).
Yule's first attempt is something like a: "fraction of pauperism which is reduced
by aid'', mimicking Galton.

This, clearly, did not satisfy Yule and we see this in Yule (1907).
Yule was the first to explicitly look for a new, general,
interpretation of regression parameters, independent on the specific
context where the regression was used and only dependent on the properties
of regression.

A priori, there was no obvious reason for such a general interpretation to
exist in any significant an direct way.
In fact, it took Yule eight years to arrive at a positive result:
from Yule (1899) to Yule (1907). The fundamental new tool introduced
by Yule is the partial regression theorem. Once found this result,
Yule clearly (if in his usual understated way) states its central role in interpreting regression results.

Yule's work on regression betas is a good example of the rapid evolution
of Statistics at the turn of the 20th century. A unified theory of
inference was being assembled out of a host of ad hoc methods
in a multitude of fields. While, as a rule, each of these methods, had a specific
justification in its field of origin, the use of these methods outside
their original setup required a new approach to their justification.
Statistical methods began to be justified in general statistical terms,
not restricted to the fields where such methods had been invented.

\section{Partial regression theorem.}

The main new result in Yule (1907), from the interpretative point
of view, is the partial regression theorem (PRT). 
The role of the PRT in the  interpretation of regression betas,
is to show that $\beta_{j,Y|X}$ is the parameter in a particular
regression between two variables.  
The point is to specify which are the regressor and the dependent variable
in this particular univariate regression. 

A note is of the order. Yule and most of his contemporaries do not use a different notation for 
population an sampling quantities. As noticed by Aldrich (1998) and (2005) the modern notation is slowly introduced in a following period
and becomes dominant around 1930. For simplicity, in this paper we use a population notation while 
 Yule's results are to be understood as sampling results. In the OLS setting we consider, the only difference between sampling and population results, under standard hypotheses,  is the use of means as opposed to expected values. 
 
\subsection{Some notation.}
Let 
\[
E(Y|X_{1},...,X_{k})=\sum_{j=1}^{k}X_{j}\beta_{j,Y|X}=X\beta_{Y|X}
\]

Without loss of generality, in order to simplify formulas, we suppose
all (marginal) expected values of all variables to be equal to 0.

The notation $\beta_{j,Y|X}$ for the coefficient of the $j-th$ column
of $X$ in the regression of $Y$ on $X$ is a version of the notation
first introduced in Yule (1907).

\subsection{Two versions of the PRT.}
A first version of the partial regression theorem states that $\beta_{j,Y|X}$ is identical to
the parameter of the univariate regression of $Y$ on $\Delta_{X_{j}|X_{-j}}=X_{j}-E(X_{j}|X_{-j})$. 

This is very general in terms of the form of $E(X_{j}|X_{-j})$ however,
following the general spirit of this paper, we shall suppose that
either we are considering linear regressions or linear approximations
of regressions, hence $\Delta_{X_{j}|X_{-j}}=X_{j}-X_{-j}\beta_{X_{j}|X_{-j}}$

In words: $\Delta_{X_{j}|X_{-j}}$ is the "part of $X_{j}$'' which
is uncorrelated with $X_{-j}$.

A second version of the PRT is: $\beta_{j,Y|X}$ is identical to the parameter of the regression
of $Y-X_{-j}\beta_{Y|X_{-j}}$ on $\Delta_{X_{j}|X_{-j}}$.

To each of these two univariate regressions corresponds an $R^{2}$. 
In the case of $Y$ on $X_{j}-X_{-j}\beta_{Y|X_{-j}}$ this is the
"semi partial $R^{2}$''.
In the case of $Y-X_{-j}\beta_{Y|X_{-j}}$ on $X_{j}-X_{-j}\beta_{Y|X_{-j}}$
this is the "partial $R^{2}$''. 

The partial correlation coefficient dates back at least to Pearson
(1896). The semi partial correlation coefficients is implied in Yule
(1907). The first use of the term "semi partial correlation'' is,
to the best of our knowledge, in Dunlap and Cureton (1930) who base
their analysis on the 1919 edition of Yule (1911).
Both these correlation coefficients are useful to measure the "relevance''
(at least in terms of fit) of a regressor taking into account the presence of other regressors.

The computation of $\Delta_{X_{j}|X_{-j}}$ is usually called "partialling
out'' $X_{-j}$ from $X_{j}$, or or "adjusting'' $X_{-j}$ for
$X_{j}$. accordingly, $X_{-j}\beta_{Y|X_{-j}}$ is sometimes called
"partial regression''. Hence the name of the theorem.

The ability of reinterpreting a multivariate regression as a set of
univariate regressions is the fundamental step in the interpretation
of a regression introduced in Yule (1907). These regressions do not involve $X_{j}$ but $\Delta_{X_{j}|X_{-j}}$. 
Given $X_{j}$, a different $X_{-j}$ implies, in general, different
values and meanings of $\beta_{j,Y|X}$. In fact, each $X_{-j}$,
in general, implies a different $E(X_{j}|X_{-j})$ and $\Delta_{X_{j}|X_{-j}}$.
The partial regression notation introduced in Yule (1907) is centered on this point. 


\section{Regression Beta: an origin story.}
We are now in the position to go into the details of our story.

As summarized above, it is in the context of Galton's "regression''
that $\beta_{j,Y|X}$ gets a first interpretation in terms of "average fraction
of a character'' passed from parents ($X)$ to offspring ($Y$).

The development of  modern theory and practice of linear regression
after Galton is mostly dependent on a series of works by Edgeworth, Karl Pearson
and Yule. This pioneering period is concluded and summarized  in Yule (1907) and in Yule
(1911). In particular Yule (1911) is a textbook which,  for almost
50 years and 14 editions up to 1958, shall have an important role
in the formation of statisticians. 

While Yule is not the originator of regression analysis, which mostly
comes from Galton and Pearson, Yule is, arguably, the first researcher
who departs from the joint Gaussian model and the first to present
multiple regression and in particular, partial regressions, in a modern
way. In comparison with modern treatises, the definitive version of
Yule's treatment of regression: Yule (1911), lacks most of the inferential
treatment introduced by Fisher in the nineteen twenties (see Aldrich
(2005)) and the use of modern matrix notation which shall enter the
field with Aitken (1935). 

Yule's book was so well known that since the twenties and thirties
of the past century its results were used without explicit quotation.
In some case this created the wrong impression that some of Yule's
results were, actually, due to other researchers. In what follows
we shall briefly discuss the attribution of the partial regression theorem.
The result is in Yule (1907) and is reproduced in Yule (1911),
but is now frequently quoted as "Frisch-Waugh theorem'' from a
paper written in 1933 (Frisch and Waugh (1933)).

Joint with Pearson and Yule, the third main figure in the early development
of regression after Galton is C. Y. Edgeworth. Edgeworth writes many 
important papers on technical points of regression computations.
While at the origin of the formulas for multiple regression, he seems
not interested in actually using regression. We found a single example
based on Galton's data.

Edgeworth, as most authors before Yule, always works in the joint Gaussian hypothesis. $E(Y|X_{1},X_{2})$
is defined as the value of $Y$ which maximizes the joint density
of $Y,X_{1},X_{2}$ for given values of $X_{1}$ and $X_{2}$. In other
words: the regression is the best forecast for given values of the
other variables. 

Edgeworth's approach, as Pearson's, is a strong clue to the fact that,
in Galton's circle, regression was intended more as a tool for producing
"best fit'' or "best forecast'' than as a tool to estimate the
parameters of a preexisting model with observations "disturbed''
by noise, as was the case for Gauss. 

 The strict connection with the Gaussian distribution
shall be abandoned, with much resistance by Pearson, only thru the
work of Yule and, later,  of R. A. Fisher (see Aldrich (2005)).

No specific interpretation of the betas is given by Edgeworth, hence,
while important for the story of regression, Edgeworth's work does
not contribute to the story of regression interpretation. 

Our main sources shall be: Pearson (1896) and Yule (1899),
(1907) and (1911).

\subsection{Pearson (1896). }

Pearson (1896) is, arguably,  Karl Pearson's  most celebrated paper
about regression. In it, he summarizes years of research and gives
an extensive treatment of multiple regression.
Pearson introduces multiple regression as a generalization of what
Galton did studying heredity. Pearson offers an interpretation of the
betas in a multiple regression which is strictly connected with Galton's
interpretation. 

In section 7 of his 1896 paper, Pearson studies the joint Gaussian
distribution of the deviations from the means of the "sizes of three
organs''\footnote{These could be different organs.}, one in the
offspring and one in each of the two parents. Then he proceeds to
compute the regression function of the offspring organ size deviations
from the mean on those of the parents. This is in full agreement with
Galton's analysis except for the fact of using the two parental charachters separatedly, in a true multivariate regression. 

In fact,  Pearson (1896) (pp. 275,
288 and ff.) criticizes Galton's "mid parent'' approach. Pearson's
idea is that it is possible to interpret the measure of a given character
in an offspring, as a linear combination of measures of characters
in the parents where the two characters  are not reduced to "mid-parent".

Pearson does not consider regression as a model for heredity, just a useful
statistical summary. This is very clearly stated in
 Pearson (1896) (p. 255):
\begin{quotation}
In the first place, we must definitely free our minds, in the present
state of our knowledge of the mechanism of inheritance and reproduction,
of any hope of reaching a mathematical relation expressing the degree
of correlation between individual parent and individual offspring.
The causes in any individual case of inheritance are far too complex
to admit of exact treatment;
\end{quotation}
In Pearson (1903) (p. 217):
\begin{quotation}
The above is in no sense a biological theory, it is based on no data whatever
except the actual statistics; it is merely a convenient statistical method of
expressing the observed facts. If the facts are there it expresses them up to
a certain point,-the most probable or most frequent value of the individual
given his relatives
\end{quotation}
As discussed above, we have a case of regression without structural model, a best fit use of regression. Quite far from the original Legendre-Gauss setup.

In section 8 of Pearson (1896) an interpretation of a regression with
two regressors is given, probably for the first time.
(The notation in what follows is ours).

Pearson starts from the hypothesis of non correlated $X_{1}$ and
$X_{2}$. This is called "non assortative'' mating: mates are not
chosen on the basis of the characters under analysis.
In this case, $\beta_{j,Y|X_{1},X_{2}}=\beta_{Y|X_{j}}$for $j=1,2$. 

Pearson goes on allowing for the possibility
of assortative mating and, in this way, introduces correlation between
the regressors. This, however, only in a very particular case: $\beta_{Y|X_2}=0$.
Let us quote the paper from page 289: 
\begin{quotation}
Let us suppose, what is not improbable, that there is a first organ,
say in the father, which has no sensible correlation with a second
organ in the offspring, but that the latter organ in the mother is
closely correlated by assortative mating with the first organ in the
father.
\end{quotation}
In our notation and in the case of two regressors
we have:
\[
\beta_{1,Y|X_{1},X_{2}}=\frac{\beta_{Y|X_{1}}-\beta_{Y|X_{2}}\beta_{X_{2}|X_{1}}}{1-\beta_{X_{2}|X_{1}}\beta_{X_{1}|X_{2}}}=\frac{\beta_{Y|X_{1}}-\beta_{Y|X_{2}}\beta_{X_{2}|X_{1}}}{1-\rho_{X_{1},X_{2}}^{2}}
\]
Analogously for $\beta_{2,Y|X_{1},X_{2}}$.
Hence, under Pearson's hypotheses we have:
\[
E(Y|X_{1},X_{2})=\frac{\beta_{Y|X_{1}}}{1-\rho_{X_{1},X_{2}}^{2}}X_{1}-\frac{\beta_{Y|X_{1}}\beta_{X_{1}|X_{2}}}{1-\rho_{X_{1},X_{2}}^{2}}X_{2}
\]
 (here 1 is the mother and 2 is the father. Recall that, due to the
joint Gaussian hypothesis assumed by Pearson, all regressions are
linear).

Pearson concludes (implicitly supposing all correlations to be positive):
\begin{quotation}
This shows us that the possession in any exceptional degree of the
first organ by the father will actually reduce the amount of the second
organ which the offspring inherits from the mother. 
\end{quotation}
Here we notice that, while $\beta_{1,Y|X_{1},X_{2}}$ is still seen
as "a fraction of the character of the mother passed to the offspring'',
$\beta_{2,Y|X_{1},X_{2}}$ is seen as the measure of an attenuation
of this "character transmission'' depending on the intensity of
the father's (different) character.

How can we evaluate such a reading? Fortunately, we can answer this question on the basis of an example by Pearson which we fully quote. Thesymbols indicating the variables are
changed to adapt these to our notation.
\begin{quotation}
Suppose the problem
to be the inheritance of artistic sense from the mother and ($Y,X_{1}$)
be measures of the deviations of this sense in son and mother from
the normal. Suppose further that $X_{2}$ be a measure of the father's
physique, say his girth of chest. Now it is conceivable that artistic
sense in the mother may be closely correlated with physique in father.
If now we deal with artistic sense of the son as related to physique
in father and artistic sense in mother, we conclude that exceptional
physique in the father will reduce the exceptional artistic sense
which the son inherits from his mother. Similarly, the exceptional
physique which the son would inherit from his father would be reduced
by exceptional artistic sense in his mother. It will be noted that
these results have no relation whatever to the coexistence or not
of artistic sense with physique in the father or the mother. They
depend entirely on the influence of assortative mating. It is remarkable
that, given mothers of high artistic sense, then this will be handed
down in a greater degree to those offspring whose fathers have a physique
below the average, than to those of fathers who have a physique above
the average. 
\end{quotation}
We better understand this if we recall that, under Pearson's hypotheses: 
\[
\frac{\beta_{Y|X_{1}}}{1-\rho_{X_{1},X_{2}}^{2}}=\beta_{1,Y|X_{1},X_{2}}
\]
We then have: 
\[
E(Y|X_{1},X_{2})=\beta_{1,Y|X_{1},X_{2}}X_{1}-\beta_{1,Y|X_{1},X_{2}}\beta_{X_{1}|X_{2}}X_{2}=
\]
\[
=\beta_{1,Y|X_{1},X_{2}}(X_{1}-\beta_{X_{1}|X_{2}}X_{2})=\beta_{1,Y|X_{1},X_{2}}\Delta_{X_{1}|X_{2}}
\]
In the light of this, Pearson's statements, which may at first appear
vague, become very precise and we can interpret them as a
reading of $\beta_{1,Y|X_{1},X_{2}}$ based on a simple case of the
partial regression theorem, joint with a Galtonian "fraction of
character transmitted'' interpretation of the parameter. 

As we know, $\beta_{1,Y|X_{1},X_{2}}$ does not apply to $X_{1}$
but to 
\[
(X_{1}-\beta_{X_{1}|X_{2}}X_{2})=\Delta_{X_{1}|X_{2}}
\]
We understand what Pearson means when he says that a given positive $X_{1}$
"exceptional artistic sense in the mother'' does not translate
into a proportional conditional expectation of "artistic sense in
the offspring'', this because it shall be on average accompanied
by an "exceptional physique in the father'' so that, for instance,
for a positive $X_{1}$ (and a positive $X_{2}$), we have $\Delta_{X_{1}|X_{2}}<X_{1}$.

This is further confirmed by the wording used by Pearson in the numerical
example following the statement.

In view of our understanding of regression, Pearson's ``attenuation'' idea: ``exceptional
physique in the father will reduce the exceptional artistic sense
which the son inherits from his mother...'', may seem incorrect:  the introduction in a regression of a variable
as $X_{2}$ uncorrelated with $Y$ but correlated with $X_{1}$ cannot
"crowd out'' $X_{1}$. In fact this is not Pearson's point.

Since, $X_{2}$ is uncorrelated with $Y$ , while the variance of
$(X_{1}-\beta_{X_{1}|X_{2}}X_{2})$ is smaller that that of $X_{1}$,
the covariance with $Y$ is the same. What the partial regression
does is, in this case, to take out from $X_{1}$ "useless'' variance,
uncorrelated with $Y$.

This implies  
\[\beta_{1,Y|X_{1},X_{2}}=\beta_{Y|X_{1}}/(1-\rho_{X_{1},X_{2}})>\beta_{Y|X_{1}}\]
We have a bigger multiple of a smaller quantity and, in fact 
\[
V(\beta_{1,Y|X_{1},X_{2}}(X_{1}-\beta_{X_{1}|X_{2}}X_{2}))=\beta_{Y|X_{1}}^{2}V(X_{1})/(1-\rho_{X_{1},X_{2}})\ge\beta_{Y|X_{1}}^{2}V(X_{1})
\]
In short: Pearson's argument has to do with the values of the parameters and does not extend to "explained variance''.


From this we also see immediate problems in extending the "fraction
of character'' interpretation of Galton to a multiple
regression.However, in order to fully understand Pearson's ideas on regression interpretation in the bivariate case, we should
have an example where the hypothesis $\beta_{2,Y|X_{1},X_{2}}=0$
is not held, as this hypothesis implies that
\[
E(Y|X_{1},X_{2})=\beta_{1,Y|X_{1},X_{2}}\Delta_{X_{1}|X_{2}}
\]
So that, the interpretation of $\beta_{1,Y|X_{1},X_{2}}$ is the interpretation
of the full regression.

In fact, in the rest of the paper Pearson does consider, in numerical
terms, the general case but does not give a detailed interpretation
of the results similar to what he does in the example we discussed.

This curiosity cannot be satisfied: while Pearson discussed regression in many following  papers, he never, at the best of our knowledge,
considered the general interpretation of parameters again.

\subsection{Yule (1899) and (1907): the partial regression theorem.}
The next paper relevant for our research is Yule (1899). In this paper
Yule applies the general results from his and Pearson's previous papers to the study of the "causes
of changes in pauperism''. The origins of the problem studied by
Yule and the reason why he was induced to dedicate several papers
to this are described in detail in Stigler (1986) and
shall not be discussed here.

Yule's is among the first applications of the "new'' regression,
as opposed to simple least squares, outside Genetics. With the proviso
that many, in those years, considered pauperism a genetic trait.

Yule (1899) has its origins in five previous papers: Yule (1895),
(1896) and (1896b) on pauperism and relief and Yule (1897) and (1897b)
containing most of Yule's early theory work on linear regression.

It is very important to notice that Yule is, in fact, studying what today would be called a ``policy'' problem.
The problem is: is it better, for reducing pauperism, to help paupers with is or with out relief. Here ``in relief'' means
relief in instituzionalized setting, as in workhouses.
To all extent, Yule is studying ``observational data'' to solve a ''treatment'' problem. This makes betas interpretation central to his analysis.

Yule (1899, p. 251) clearly states, in words which we could find identical in modern textbooks,  the reason why conditioning on
more than one variable is relevant: in order to avoid...
\begin{quotation}
...the disadvantage of the possibility of a double interpretation,
as mentioned above: the association of the changes of pauperism with
changes in proportion of out-relief might be ascribed either to a
direct action of the latter on the former, or to a common association
of both with economic and social changes.
\end{quotation}
However, Yule does not jump directly to the use of linear (in the
parameters and variables) regression. While still not in posses of
the PRT (we shall see Yule's comment to this exact point in Yule (1907)),
  Yule (1897, p.833) states:
\begin{quotation}
$\rho_{1,2|3}$ is the coefficient of correlation for any group of
$X_{1},X_{2}$'s associated with a single type of $X_{3}$'s so long
as the conditions hold that the means of all arrays are linear functions
of their types. In general correlation where the net coefficient will
change as we pass from one type [meaning: value] of $X_{3}$ to another, $\rho_{1,2|3}$
can only retain an average significance
\end{quotation}
This statement originates in Yule (1897b), a preliminary version of
Yule (1897), where linear regression is discussed beyond the Gaussian
case. In Yule (1897b) it is made clear that the validity of the first
part of the quoted statement holds in the case of joint Gaussian distribution.
in Yule (1897) the statement is slightly more general, as the condition
is "means of all arrays are linear functions of their types''.

Yule (1897) does not go further. We still need a last piece of evidence
before quoting Yule (1899). In Yule (1896, p. 615) partial correlation
coefficients are described as:
\begin{quotation}
coefficients of correlation between any two of the variables while
eliminating the effect of variations in the third (or others)
\end{quotation}
While accompanied by no proof, this almost is a statement of the PRT, at least in a particular important case.
As Yule shall write in Yule (1907), what still is lacking is the ability of computing the
partial correlation coefficient  as the simple correlation
between two specific variables built out of $Y$ and $X$.

This preparatory work takes us to the first complete attempt by Yule
at interpreting a parameter in a multiple regression: Yule (1899,
p. 251), emphasis is ours. 
\begin{quotation}
Then suppose a [..] regression equation to be formed
from these data, in the way described in my previous paper {[}Yule
(1897){]}, first between the changes in pauperism and changes in proportion
of out-relief only. This equation would be of the form

change in pauperism = A + B x (change in proportion of out-relief) 

where A and B are constants (numbers)

This equation would suffer from the disadvantage of the possibility
of a double interpretation, as mentioned above: the association of
the changes of pauperism with changes in proportion of out-relief
might be ascribed either to a direct action of the latter on the former,
or to a common association of both with economic and social changes.
But now let all the other variables tabulated be brought into the
equation, it will then be of the form

change in pauperism= a + b x (change in proportion of out-relief)
+ c x (change in age distribution) + d x ... + e x ... (changes in
other economic, social, and moral factors)

Any double interpretation is now-very largely at all events excluded.
It cannot be argued that the changes in pauperism and out-relief are
both due to the changes in age distribution, \emph{for that has been
separately allowed for }in the third term on the right; 

b x (change in proportion of out-relief) 

gives the change due to this factor \emph{when all the others are
kept constant}. 
\end{quotation}
With our knowledge of Yule's two previous papers, we  understand
the meaning of this statement. The "allowed for'' reprises Yule
(1896) while the "kept constant'' statement is not to be considered
as at odds with "allowed for", but can be
correctly understood in the light of the above quotation 
of Yule (1897).

However, this statement is still vague and imprecise. 
Yule is conscious of this this is testified by the eight years he dedicates to solve this problem. The final results of this work are presented in his 1907 paper.
 
Yule (1907)  contains a substantially complete theory of multiple
linear regression. The approach systematizes past intuitions of both Pearson and Yule but is new in its extent and precision. It is centered on the PRT notion of "partial out" a variable from other variables. This central role given to PRT goes to the point of introducing a new "partial out'' notation for the betas so succesfull to  be  the one still in use today.

The partial
regression theorem is introduced as the result which allows to express any coefficient in a multiple
regression as a coefficient in a simple regression. 
The version of PRT in Yule (1907, p.184), is
(using our notation):
\begin{quotation}
$\beta_{j,Y|X}$ may be regarded, quite generally and without any
reference to the form of the frequency distribution, as the regression
of $Y-X_{-j}\beta_{Y|X_{-j}}$ on $X_{j}-X_{-j}\beta_{X_{j}|X_{-j}}$.
\end{quotation}
While the PRT shares clear relationships with general orthogonalization
methods, known since at least Gauss, it is Yule, to our best knowledge,
the first to prove PRT in a regression context and, most relevant,
the first to understand the power of the result for interpreting regressions.

The following
quotation from Yule (1907, p.184) testifies Yule's position on this point:
\begin{quotation}
This result is of some importance as regards the interpretation of
partial correlations and regressions. In the case of normal correlation
there is no difficulty in assigning a meaning to these constants,
as the regression is strictly linear, and the partial correlations
and regressions are the same for all types of the variables. But in
the general case this is not so, and although I showed, in a previous
discussion of the question, that the values assigned to the partial
regressions on the assumption of normal correlation are the 
"least square" values and, consequently, that the partial
correlation retains an "average significance",
I could not prove that it remains an actual correlation between determinate
variables. The above theorem completes the work in this respect''.
\end{quotation}
Yule  pulls together and connects the many strains  already present in
previous research. The case of normal correlation is the case
of regressions which are both linear in the parameters and in the
variables, so that the partial regression coefficients are the same
for all the values (types) of the other variables. Hence, in principle, we could compute these coefficients after selecting 
those data points which have given values of the other variables and then running an univariate regression in the selected sample.

In general this
is not true (for instance it is not true if the regression is linear
in the parameters but not in the variables) and we have an "average''
meaning as the single partial correlation shall be an average of the
different partial correlations for the different values of the other
variables. But this is now solved as the PRT shows that each partial
correlation (o regression) coefficient IS in ANY case the correlation (regression)
coefficient between two specific variables: the residuals of $Y$ and $X_{j}$ regressed on $X_{-j}$.

This ``completes the work": thanks to this result, a complete interpretation of regression betas is now possible.

After the statement and proof of this result and once explained its
importance for the interpretation of a multiple regression, Yule proceeds
(Yule (1907) section 16) to analyze with this tool the pauperism dataset already
considered in Yule (1899). 

\subsection{Yule (1911), a textbook and its influence.}

In 1911 Yule publishes a textbook: Yule (1911). The multiple regression chapter is an
extended version of Yule (1907). 
The influence of Yule's textbook in the first half of the XXth century cannot be overstated.
This made Yule's contribution widely known, but, as a negative consequence, the Author behind it was often forgotten.

In his 1955 obituary of Yule, Sir. Maurice Kendall, who collaborated
with Yule in the later editions of Yule (1911), wrote:
\begin{quotation}
The value of some of his contributions has been lost to view in sheer
virtue of their success; for example, his work on correlation and
regression is now such standard practice that only a student of history
would consult the original papers
\end{quotation}
A clear and fairly well known testimony of this is in Frisch and Waugh
(1933) (see, for instance, Aldrich (1998)). 
The partial regression theorem is today sometimes called the "Frisch
Waugh theorem'' as it can be found in Frisch and Waugh (1933). 
Yule's work is not quoted in Frisch and Waugh (1933), which is not
surprising, being Yule (1911) a widely known textbook at the time.\footnote{While quite well known a fact, at least among historians of statistics (see e.g. Aldrich (1998)), the precedence of Yule on, for instance, Frisch and Waugh, is, from time to time, ``rediscovered''. This is not surprising if we think that the partial regression theorem itself has been ``rediscovered'' several times. At least up to the 1950s.}  

We know that the discussion of partial regressions
contained in Yule (1907) and Yule (1911) was well known to Frisch.
In fact, it is quoted in Frisch and Mudgett (1931) (pag. 389) to amend  Yule's formulas misquoted by another Author. The exact terms
are "the formulae for multiple and partial correlation, as given
by Yule'', again with no reference to a specific book or paper, which
testifies for the widespread knowledge and acceptance of Yule's setup
for regression theory among Statisticians.\footnote{The formulas discussed by Frisch and Mudgett (1931) as 7.1 and 7.3
on page 389 are formulas 17 on pag. 189 and 14 on pag. 186 of Yule
(1907). In Yule (1907) the statement of the partial regression theorem
is in formula 7 pag. 184. The same formulas are present in Yule (1911).}

Frisch and Waugh (1933), while
not quoting Yule, apply  the partial regression notation introduced
in Yule (1907). Moreover, the discussion about
formula 4.9 of Frisch and Waugh (1933) is a paraphrase of the discussion
about the equivalent formula 8 pag. 184 of Yule (1907) and formula
7 pag. 236 of Yule (1911).\footnote{Formulas and page numbers are from the 1919 edition of Yule (1911).}

As a further clue of the importance of Yule (1911), consider section
32 ("partial correlation'') of another influential textbook: Fisher
(1950, first edition 1925). In this section the reasons and the methods
for "partialling out'' variables are discussed with somewhat greater
rhetorical flourish than Yule (1907) and Yule (1911) but exactly along
the same arguments.
Again, while Yule (1911) is quoted elsewhere by Fisher (from the 1915
edition of Yule's book), this is not the case in section 32 of Fisher's book.

Yule (1911) quickly becomes well known abroad. Not only in the USA, in Germany and in the Scandinavian countries. In Italy, Giulio Mortara
(1915) publishes: "On the measure of dependence
in collective phenomena from empirical variables'' (in Italian) a keen and detailed analysis of statistical dependence, where Yule (1911) is, with singular detail w.r.t. contemporary literature, repeatedly quoted, the partial regression
theorem is used (sec. 33 pag. 39) and alternative measures of dependence
are discussed.

Mortara's precise quotations, perhaps, avoided the partial regression theorem to be nicknamed ``Mortara's theorem'' 18 years before the similar but less bibliographically completete Frisch and Waugh paper.

Overall, after Yule (1911) the stage seems quite well set for a widespread
diffusion of the  "partialling out'', interpretation of
multiple regression coefficients.
This did not happen, at least not in full. In the Introduction of this paper we did see that what can be found today, 
in books and papers, is a mix of several interpretations. In fact, few are the cases where the PRT based interpretation is the only one  presented.\footnote{A relevant example of this is the
section about  Regression in the Reference Manual on Scientific Evidence USAS (2011, p. 336).} 

Actually, it is again  in Yule (1911) that we can find the first example of this "eclectic" attitude.

In the introduction to the partial regression chapter, Yule (1911, p. 231 of 1919 edition)
 Yule states that partial correlation coefficients
are...
\begin{quotation}
...giving the correlation between $X_{1}$ and $X_{2}$ or other pair
of variables \emph{when the remaining variables $X_{3},...,X_{n}$,
are kept constant} {[}the emphasis is the Author's{]}, or\emph{ when
changes in these variables are corrected or allowed for} {[}the emphasis
is ours{]}, so far as this may be done with a linear equation
\end{quotation}
This is a reprise of the 1899 statement. 
Following this statement the analysis  in Yule (1911) is completely based on the PRT and never resorts to "keep constant" or "controlling" statements.

If we now go back to the quotation from  Gujaraty (2009) with which we began this paper we see how what we find in 2009 is an almost exact quotation of Yule (1911).

The historical perspective followed here, understandably absent from textbooks and papers, can be useful to better qualify these statements. 


After 1911 Yule mostly worked in fields different than regression.  In 1936, many years after his last paper on regression, Yule goes back to the topic. Not surprisingly,  he does this to criticize what, by then, had become a common misinterpretation of regression betas.  Yule (1936) is a critique of the "partial derivative" interpretation of the betas which had evolved out  the early "keep constant" idea.

\section{Conclusions. Suggestions for teaching regression.}

The search for an interpretation of regression coefficient independent of the existence of an underlying "structural model"  followed a long and rather complex path. We can now perhaps better see how several ideas which we today find in books and papers, did progressively develop at the end of nineteenth century and how these got clarification in Yule's work.

We can derive from this some possible suggestions for the  teaching of the use and interpretation of regression models.

As a rule, in teaching linear regression, the linear model is first
introduced, then weak-OLS hypotheses are added and OLS estimation and its properties are then discussed.
The discussion about how to read the results is introduced after a
presentation of the basic estimation theory.

Alternatively, linear regression as a best fit to $Y$  should be
introduced first. The problem of when a best fit to $Y$ implies a good estimate of a  $\beta$ in a structural model, if this exists, should then be discussed. 
The basic assumption $E(\epsilon|X)=0$ should be compared with alternatives, e.g. those originating instrumental variables. 

Interpretation of regression parameters, considered as known, should precede estimation theory. Conditions under which approximate ideas like "keep constant" and "control for" are useful can be made precise in the light of PRT.

This small twist in the syllabus could help the student toward a better
understanding of the relationship between linear models and linear
regression. The possible addition of some history of the subject as it developed in the study of heredity, maybe along
the lines considered here, could be of further help.


\begin{thebibliography}{}

\bibitem[Aldrich (1998)]{key-1c} Aldrich, J. (1998) Doing Least Squares:
Perspectives from Gauss and Yule. \emph{International Statistical
Review 66(1),} 61-81.

\bibitem[Aldrich (2005)]{key-1d-1} Aldrich, J. (2005) Fisher and
Regression. \emph{Statistical Science 20(4)}, 401-417 

\bibitem[Box (1966)]{key-42}Box, G. E. P. (1966). Use and Abuse of
Regression. \emph{Technometrics 8}(4), 625-629. 



\bibitem[Browne (2002)]{key-41b}Browne, J. (2002). \emph{Charles
Darwin---The Power of Place}. London: Jonathon Cape.  




\bibitem[Dunlap and Cureton (1930)]{key-36}Dunlap, J. W. and E. E.
Cureton (1930). On the analysis of causation. \emph{Journal of Educational
Psychology 21}, 657-679. 

 
\bibitem[Fisher (1950)]{key-34}Fisher, R. A. (1950). \emph{Statistical
Methods for Research Workers}. Oliver and Boyd (ed. X, first edition
in 1925). 


\bibitem[Frisch and Mudgett (1931)]{key-32}Frisch, R. and B. D. Mudgett
(1931). Statistical Correlation and the Theory of Cluster Types. \emph{Journal
of the American Statistical Association 26}(176), 375-392. 

\bibitem[Frisch and Waugh (1933)]{key-31}Frisch, R. and F. V. Waugh
(1933). Partial Time Regressions as Compared with Individual Trends.
\emph{Econometrica 1}(4): 387--401. 

\bibitem[Gauss-Stewart (1995)]{key-111} Gauss, C. F. and G. W Stewart
(translator) (1995). \emph{Translation of Gauss (1821): Theoria Combinationis Observationum Erroribus
Minimis Obnoxiae}. SIAM.

\bibitem[Griffin (1931)]{key-30}Griffin, H. D. (1931). On Partial
Correlation VS. Partial Regression for Obtaining the Multiple Regression
Equations. \emph{The Journal of Educational Psychology 22}, 35-44.


\bibitem[Gujarati and Porter (2009)]{key-31-1} Gujarati, D. N. and
D. C. Porter (2009). \emph{Basic Econometrics.} 5th ed. McGraw-Hill. 



\bibitem[Legendre (1805)]{key-123}Legendre, A. M. (1805). \emph{Nouvelle
Methodes pour la Determination des Orbites des Cometes}. Firmin Didot,
Paris

\bibitem[Mortara (1915)]{key-1d}Mortara, G. (1915). Sulla misura
della Dipendenza di Fenomeni Collettivi da Variabili Empiriche. \emph{Giornale
degli economisti e Rivista di Statistica, Serie terza}, 50, (2), 1-42. 

\bibitem[Mosteller and Tukey (1977)]{key-33}Mosteller, F. and J.
W. Tukey (1977). \emph{Data Analysis and Regression a Second Course
in Statistics}. Addison-Wesley. 

\bibitem[Pearson (1896)]{key-2}Pearson, K. (1896). Mathematical Contributions
to the Theory of Evolution. III. Regression, Heredity, and Panmixia.
\emph{Philosophical Transactions of the Royal Society of London. Series
A, Containing Papers of a Mathematical or Physical Character 187},
253-318. 

\bibitem[Pearson (1903)]{key-2b}Pearson, K.(1903). The Law of Ancestral Heredity.
\emph{Biometrika, Vol. 2, No. 2}, 211-223.


\bibitem[Pearson (1929)]{key-3}Pearson, K. (1920). Notes on the History
of Correlation. \emph{Biometrika 13}(1), 25-45. 

\bibitem[Pearson (1930)]{key-4}Pearson, K. (1930). \emph{The Life,
Letters and Labors of Francis Galton}. (Vol 3) Cambridge University
Press. 

\bibitem[Seal (1967)]{key-8}Seal, H. L. (1967). Studies in the History
of Probability and Statistics. XV The historical development of the
Gauss linear model. \emph{Biometrika 54}(1 and 2), pp. 1-24. 



\bibitem[Stigler (1986)]{key-12}Stigler, S. M. (1986). \emph{The
History of Statistics}. Harvard University Press. 

\bibitem[USAS (2011)]{key-121}United States Academy of Sciences (2011). \emph{Reference Manual on Scientific Evidence}. 3rd ed. The National Academies Press.

\bibitem[Yule (1895)]{key-25-2}Yule, G. U. (1895).On the Correlation
of total Pauperism with Proportion of Out-Relief. \emph{The Economic
Journal }5(20), 603-611.

\bibitem[Yule (1896)]{25-1-12}Yule, G. U. (1896).On the Correlation
of total Pauperism with Proportion of Out-Relief. \emph{The Economic
Journal }6(24), 613-623.

\bibitem[ Yule (1896b)]{key-25-1}Yule, G. U. (1896b). Notes on the
History of Pauperism in England and Wales from 1850, Treated by the
Method of Frequency-Curves; with an Introduction on the Method. \emph{Journal
of the Royal Statistical Society }59(2), 318-357. 

\bibitem[Yule (1897b)]{key-25-5}Yule, G. U. (1897b). On the Significance
of Bravais' Formula for Regression, \&c., in the case of Skew Correlation.
\emph{Abstracts of the Papers Communicated to the Royal Society of
London} 60(1), 477-489. 

\bibitem[Yule (1897)]{key-25}Yule, G. U. (1897). On the Theory of
Correlation. \emph{Journal of the Royal Statistical Society 60}(4),
812-854.

\bibitem[Yule (1899)]{key-24}Yule, G. U. (1899). An Investigation
into the Causes of Changes in Pauperism in England, Chiefly During
the Last Two Intercensal Decades (Part I.). \emph{Journal of the Royal
Statistical Society 62}(2). 249-295. 

\bibitem[Yule (1907)]{key-23}Yule, G. U. (1907). On the Theory of
Correlation for any Number of Variables, Treated by a New System of
Notation. \emph{Proc. R. Soc. Lond. A 79}, 182-193. 

\bibitem[Yule (1911)]{key-22}Yule, G. U. (1911). \emph{An Introduction
to the Theory of Statistics}. Charles Griffin \& Co. 

\bibitem[Yule (1936)]{key-21}Yule, G. U. (1936). On a Parallelism
between Differential Coefficients and Regression Coefficients. \emph{Journal
of the Royal Statistical Society 99}(4), 770-771. 

\end{thebibliography}
\end{document}